\documentstyle[aps,epsfig]{revtex}
\def\beqa{\begin{eqnarray}}
\def\eqar{\end{array}}
\def\beqar{\begin{array}}
\def\eqa{\end{eqnarray}}
\def\bars{\begin{eqnarray*}}
\def\ears{\end{eqnarray*}}

\def\beq{\begin{equation}}
\def\eq{\end{equation}}

\newcommand{\bt}{\begin{tabular}}
\newcommand{\et}{\end{tabular}}
\newcommand{\bd}{\begin{displaymath}}
\newcommand{\ed}{\end{displaymath}\noindent}
\newcommand{\ec}{\end{center}}
\newcommand{\bc}{\begin{center}}

\begin{document}
\title{A new determination of the Pomeron intercept in hard processes}
\author{J. G. Contreras\thanks{Departamento de F\'{\i}sica Aplicada,
 CINVESTAV--IPN, Unidad M\'erida, A. P. 73 Cordemex, 97310 M\'erida, 
 Yucat\'an, M\'exico},
 R. Peschanski\thanks{%
CEA, Service de Physique Theorique, CE-Saclay, F-91191 Gif-sur-Yvette Cedex,
France}
C. Royon\thanks{%
 Service de Physique des Particules,
CE-Saclay, F-91191 Gif-sur-Yvette Cedex, France}}
\maketitle

\begin{abstract}
A method allowing for a direct comparison of data with theoretical  predictions 
is 
proposed for forward jet 
production at HERA. It avoids the reconstruction of multi-parton contributions 
by 
expressing the 
experimental cuts directly as correction factors on the QCD forward jet 
cross-section. 
An application to the determination of the {\it effective } Pomeron intercept in 
the 
BFKL-LO parametrization from $d\sigma/dx$ data at HERA   leads to a good fit 
with a 
significantly higher {\it effective }
intercept, $\alpha_P= 1.43 \pm 0.025 (stat.) \pm 0.025 (syst.),$ than for proton 
(total 
and diffractive) structure functions. It is however less than the value of the 
pomeron
intercept using dijets with large rapidity intervals obtained at Tevatron.
We also evaluate the rapidity veto 
contribution to the higher order BFKL corrections.
The method can be 
extended to other theoretical inputs.
\end{abstract}

\bigskip

{\bf 1. Introduction}
 
\bigskip
The study of forward jets at colliders is considered as the milestone of QCD 
studies 
at high energies,
since it provides a direct way of testing the perturbative resummations of soft 
gluon 
radiation. More 
precisely, the  study of one forward jet (w.r.t. the proton) in an 
electron-proton 
collider~\cite{mu91}
seems to be a good candidate to test the energy dependence of hard QCD 
cross-sections. It is similar to the previous proposal of studying two jets 
separated by 
a large rapidity interval in hadronic 
colliders~\cite{mu86}, for which only preliminary results are 
available~\cite{D0}. This 
test is also possible in $\gamma^*$-$\gamma^*$ scattering~\cite{ro99} but here 
the 
statistics and the energy range are still insufficient to get a reliable 
determination 
of the physical parameters for hard QCD 
cross-sections.
Indeed, the 
proposed (and favored for the moment being) set-up~\cite{mu91} is to consider 
jets with 
transverse momentum $k_T$ of the  
order of the 
photon virtuality $Q$ allowing to damp the  QCD evolution as a function of $k_T$ 
(DGLAP 
evolution~\cite{al77}) in favor of the   evolution in energy at fixed $k_T$ 
(BFKL 
evolution~\cite{li77}).

Since proposal \cite{mu91} was made, a set of interesting studies have been 
performed to 
check its relevance. On the 
experimental 
ground, H1~\cite{h199} and ZEUS~\cite{ze99} have published useful results with 
appropriate cuts (to be 
displayed 
later on) at relatively small $x.$ On the theoretical ground, the general
formulation and some quantitative estimates have been performed prior to 
experiments~\cite{ba92} confirming 
the 
interest in such processes. The recent  theoretical analyses have been 
mainly based on the use of 
Monte-Carlo 
simulations, including the multi-parton cross-sections and starting from  the 
various 
frameworks in competition~\cite{po99}.
Quite a few analyses arrive at a satisfactory description of the data, taking 
into 
account the specific 
parametrizations
which are choosen. Indeed, the BFKL-based Monte-Carlo~\cite{po99} lead to quite 
satisfactory results, while those based on  DGLAP evolution meet some difficulty 
to 
describe the data\footnote{Note however, that some more refined versions of 
DGLAP 
evolution including contributions from the resolved off-mass-shell photon can 
describe 
the data~\cite{po99}.}.

However, there still remains a 
problem in the interpretation of those results. Due to the 
difficulty in handling the experimental cuts 
 without introducing in the simulation  the whole set of theoretical n-parton 
contributions 
to the cross-section, it 
appears 
difficult  to avoid the uncertainties of the 
reconstruction (with the parameters 
and 
constraints which are needed to define the scheme in practice). It seems thus 
difficult 
to determine unambiguously  genuine 
theoretical parameters defining  the cross-section 
one
is looking at. One example is the dependence in  parameters such as  infra-red 
cut-offs, 
which are not a-priori 
required in the expression of 
the 
total $d\sigma/dx$ jet cross sections. Another illustration is  the so-called 
``consistency 
constraint'' which appears very 
useful~\cite{kw96} in the expression of next-to-leading corrections to the BFKL 
formula 
coming from the n-parton contributions, but again is  not 
 expressed in terms of  
the 
$d\sigma/dx$ jet cross sections itself. In fact, it does not seem easy, in those 
schemes, to extract with some precision the value of the {\it effective} Pomeron 
intercept $\alpha_P$, i.e. the main theoretical parameter describing the 
theoretical 
energy dependence in this process. As we know, this parameter is of primordial 
importance to evaluate the amount of next-leading corrections in a BFKL 
framework~\cite{li98} and to confront its {\it effective value} with the recent 
theoretical determinations~\cite{sa99}.

We want to 
address 
this problem in a quite different way, that is on focussing on the jet cross 
section 
$d\sigma/dx$ observable itself, 
by a consistent treatment of 
the 
experimental cuts and minimizing the uncertainties for that particular 
observable. Let us remark that 
our 
approach is not intended to provide a substitution to the other methods, since 
the 
Monte-Carlo simulations have the 
great 
merit of 
making a  set of predictions for various observables. Hence, our method has to 
be 
considered as complementary
to the others and dedicated to a better  determination of the {\it effective} 
Pomeron 
intercept using the  $d\sigma/dx$ data. As we shall see, it will fix more 
precisely this
parameter, but it will leave less constrained other interesting parameters, such 
as the 
cross-section normalization.

One fruitful outcome of the method proposed in the present paper is the 
possibility of 
comparing the effective intercept with its determination in  other processes 
involving  
QCD at high energy. In 
fact, using the parameters determined from forward jets at HERA, it is possible 
to 
compare with double jet production at Tevatron following Ref.~\cite{mu86} for 
which  
preliminary 
experimental analyses have been performed~\cite{D0} and find a high value of the 
intercept ($\alpha_P = 1.7\pm.1\pm.1$ in Ref.~\cite{D0}).
It can also be confronted with the 
effective BFKL analysis  of proton structure functions at small-$x_{Bj},$ which 
give 
rather low values\footnote {Note however that taking into account the full BFKL 
formula 
may lead to higher  $\alpha_P ,$ namely  $1.2-1.3,$ see Ref.~\cite{na96}, for 
total 
structure functions and even reach $.4,$ see Ref.~\cite{mu98}, for  diffractive 
proton  
structure functions.}($\alpha_P \sim 1.1-1.21$ see Ref.~\cite{ab00}). However, 
in those 
cases, the result may be different, since non-perturbative effects related to 
the ``soft'' proton scales are expected to influence the  determination of 
parameters.

Thus, the comparison of the effective BFKL parameter $\alpha_P$ obtained for the  
forward jet production cross-section  allows for a  study of QCD at 
high energy, aiming at a better understanding of the  corrections to the 
leading-order BFKL predictions~\cite{li98,kw96}. 

The plan of our study is the following: in section {\bf 2}, we introduce the 
QCD formalism and our 
method for determining $\alpha_P .$ In the following section {\bf 3} we
determine the kinematic correction factors to the forward jet cross-section data 
on 
$d\sigma/dx$ due to the experimental cuts. In the following section {\bf 4}, we 
perform and discuss a (separately and then common) fit to H1 and ZEUS data. This 
determines the BFKL parameter $\alpha_P$ which is subsequently used in section 
{\bf 4} 
for a comparison with the  two-jet cross-section at Tevatron from the  
experimental D0 
analysis. Discussions on these results and comparison with the BFKL study of 
(total 
and diffractive) structure functions are presented in section {\bf 5} and 
conclusions and outlook in section {\bf 6}.

\bigskip

{\bf 2. Formalism}
\bigskip
 
The cross-section for forward jet production at HERA in the dipole model 
reads~\cite{ba92}:
\begin{eqnarray}
\frac{d^{(4)} \sigma}{dx dQ^2 dx_J d k_T^2 d \Phi} =
\frac{ \pi N_C \alpha^2 \alpha_S(k_T^2)}{Q^4 k_T^2} 
\ f_{eff} (x,\mu_f^2)
\ \Sigma e_Q^2 
\int_{\frac 12- i\infty}^{\frac 12+ i\infty} \frac{d \gamma}{2i \pi} 
\left( \frac{Q^2}{k_T^2} \right)^{\gamma} \times\ \ \ \ \ \ \ \ \ \ \ \ \ \ \ \ 
\ \ 
\nonumber \\
\times \ 
\exp \{\epsilon (\gamma,0) Y\} \left[ \frac{h_T(\gamma) +h_L(\gamma)}{\gamma}
(1-y) + \frac{h_T(\gamma)}{\gamma} 
\frac{y^2}{2} \right]
-\exp \{\epsilon(\gamma,1) Y\} cos 2 \Phi  \left[
\frac{h_T(\gamma)}{\gamma} 
\frac{\gamma (1-\gamma)}
{(\gamma+1)(2-\gamma)} \right]
\label{dsigma}
\end{eqnarray}
where
\begin{eqnarray}
Y &=& \ln \frac{x_J}{x} \\
\epsilon(\gamma,p)&=& \bar{\alpha} \left[ 2 \psi(1) -\psi(p+1-\gamma)
-\psi (p+\gamma) \right] \\
f_{eff} (x, \mu_f^2) &=& G(x,\mu_f^2) + \frac{4}{9} \Sigma (Q_f+ \bar{Q_f}) \\
\mu_f^2 &\sim& k_T^2\ ,
\end{eqnarray}
are, respectively, $Y$ the rapidity interval between the photon probe and the 
jet,
$\epsilon(\gamma,p)$ the BFKL kernel eigenvalues, $f_{eff}$ the effective 
structure 
function 
combination, and $\mu_f$ the corresponding factorization scale.  
The main BFKL parameter is $\bar{\alpha},$ which is the (fixed) value of the 
effective 
strong coupling constant in LO-BFKL formulae. Note that we gave for completion 
the 
full BFKL formula including the azimuthal dependence but we will stick to the  
azimuth-independent contribution with the dominant $\exp \{\epsilon (\gamma,0) 
Y\}$ 
factor.

The so-called ``impact factors''
\begin{eqnarray}
\label{defh}
\left(\begin{array}{c}
h_T \\ h_L \end{array} \right) = \frac{\alpha_S (k_T^2)}{ 3 \pi \gamma} 
\frac{(\Gamma(1 - \gamma) \Gamma(1 + \gamma))^3}{\Gamma(2 - 2\gamma) \Gamma(2 +
 2\gamma)} \frac{1}{1 - \frac{2}{3} \gamma} \left( \begin{array}{c} (1 +
 \gamma)
(1 - \frac{\gamma}{2}) \\ \gamma(1 - \gamma) \end{array} \right)\ ,
\end{eqnarray}
are obtained from the $k_T$ factorization properties~\cite{ca91} of the coupling 
of 
the BFKL amplitudes to external hard probes. The same factors can be related 
to the photon wave functions~\cite{bj71,mu98} within the equivalent context of 
the QCD 
dipole model~\cite{mu94}.

Our goal is to compare as directly as possible the theoretical parametrization 
(\ref{dsigma}) to the 
data 
which are collected in experiments~\cite{h199,ze99}. The crucial point is how to 
take 
into account the 
experimentally defined kinematic cuts listed in Table I for the reported three 
sets of 
data (two for 
H1 
with $k_T=3.5$ or $5$ GeV, and one for ZEUS). 
\vskip 1.cm
\begin{center}
\begin{tabular}{|c|c|} \hline
  H1 cuts & ZEUS cuts\\
\hline\hline
$E^{'}_{e}>$ 11 GeV & $E^{'}_{e}>$ 10 GeV \\
160 $\le \theta '_e \le 173$ deg. & ~ \\
$y>$0.1 & $y>$0.1 \\
7$\le \theta_{jet} \le$20 deg. & $ \theta_{jet} \ge$8.5 deg. \\
$k_{Tjet} \ge$ 3.5 or 5 GeV & $k_{Tjet} \ge$ 5 GeV \\
$x_{jet} >$0.035 & $x_{jet} >$0.036 \\
$0.5 < \frac{k_T^2}{Q^2} < 2$ & $0.5 < \frac{k_T^2}{Q^2} < 2$ \\
$10^{-4} < x < 4. ~10^{-3}$  & $4.5 ~10^{-4} < x < 4.5 ~10^{-2}$ \\ \hline
\end{tabular}
\end{center}
\vskip .5cm
\begin{center}
{ Table I- Experimental cuts (H1/ZEUS)}
\end{center}
\bigskip\bigskip

The main problem to solve is to investigate the effect of these cuts on the 
determination of the 
integration variables leading to a prediction for $d\sigma/dx$ from the given 
theoretical formula for 
$d^{(4)} \sigma$ as given in formula (\ref{dsigma}). The effect is expected to 
appear as 
bin-per-bin {\it 
correction factors} to be multiplied to the theoretical cross-sections for 
average 
values of the 
kinematic variables for a given $x$-bin before comparing to data (e.g. fitting 
the 
cross-sections).

The idea of our method is threefold:  {\bf i)}  for each $x$-bin, determining 
the 
average values of $x$, $Q^2$, $E_J$, $k_T$ 
 from a known and reliable Monte-Carlo simulation of the cross-sections. For 
this sake, 
we use the {\it 
Ariadne} Monte-Carlo programme~\cite{ariadne}; {\bf ii)} 
choosing a set of integration variables 
over $d^{(4)}\sigma$ in (\ref{dsigma}) in such a way to match closely the 
experimental 
cuts and minimize the variation of the cross-sections over the bin size; {\bf 
iii)}
fixing the correction factors due to the experimental cuts for each $x$-bin, by 
a 
random simulation of the kinematic constraints  with no dynamical input. 

The point {\bf i)} proposed already in~\cite{co99} allows a determination of 
which 
average values of 
the 
kinematic variables have to be taken in the theoretical formula (\ref{dsigma}) 
for 
each experimental
$x$-bin. The point {\bf ii)} comes from the crucial requirement   to minimize 
the 
variation (over the 
$x$-bin) of the   variables to be retained for the integration. Indeed, since 
the 
integration procedure 
multiplies the central value of the integrand by the size of the integration 
bins, it is 
compulsory to 
choose adequate variables which lead to a smooth dependence of the integrand and 
of the 
effect of the 
kinematic cuts.

This double stringent requirement can be solved for the forward jet 
$d\sigma/dx.$ For 
this sake we 
choose  
\begin{eqnarray}
\frac{d \sigma}{dx} = \int \left[Q^6 \frac{d^{(4)} \sigma}{dx dQ^2 dx_J d k_T^2 
d 
\Phi}\right] \times 
\Delta \left(\frac {1} {Q^2} \right)\Delta x_J \Delta \left(\frac {k_T^2} 
{Q^2}\right) 
\Delta \Phi \ .
\label{stringe}
\end{eqnarray}
The property of this non-trivial choice is the following. The integration 
variables are 
choosen in such 
a 
way that the expression in the square brackets $[Q^6 ...]$ in (\ref{stringe})
is  dependent  on the ratio $\frac {k_T^2} {Q^2}$ and not on each scale 
separately. Looking at the experimental cuts (see Table I), it becomes  clear 
that the 
choice of this
scale-invariant integrand minimizes the variation of the observable on the 
bin, while each 
scale  
$ {k_T^2}$ and $ {Q^2}$ presents large variations and thus would generate large 
integration errors. 
Indeed, various numerical studies we have performed have demonstrated that it 
was a {\it 
sine qua non}    
 stability condition for the fits. The overall azimuthal integration ($\Delta 
\Phi=2\pi)$ 
cancels the second term in (\ref{dsigma}).

\bigskip

{\bf 3. Correction Factors}

\bigskip
 
 The experimental correction factors have been determined using
a toy Monte-Carlo designed as follows. We generate flat distributions in the 
variables 
$k_T^2/
Q^2$, $1/ Q^2$,
$x_J,$ using reference intervals   which include the whole of the experimental 
phase-space (the $\Phi$ 
variable is not used in the generation since all the cross-section measurements
are $\phi$ independent). In practice, we get the correction factors by counting 
the 
numbers of events 
which fulfill
the experimental cuts given in Table {\bf I} for each $x$-bin. The correction 
factor is
obtained by the ratio to the number of events which pass the experimental cuts  
and the 
kinematic
constraints, and the number of events which fullfil only the kinematic 
constraints,i.e.
the so-called reference bin.

\begin{center}
\begin{tabular}{|c|c|c|c|c|c|} \hline
 x & $\sigma$ & $Q^2$ & $E_{jet}$ & $k_T$ & Corr. Factor (.$10^{-3}$)\\
\hline\hline
 0.00036 & 202.5 & 13.9 & 32.6 & 4.5 & 0.270 \\
 0.00073 & 342. & 21.5 & 34.4 & 5.0 & 0.993 \\
 0.0012 & 224. & 26.9 & 36.9 & 5.5 & 1.14 \\
 0.0017 & 138. & 31.4 & 38.1 & 5.8 & 1.11 \\
 0.0024 & 67. & 38.1 & 38.8 & 6.3 & 0.921 \\
 0.0035 & 32. & 47.0 & 37.9 & 6.9 & 0.711 \\
\hline\hline
\end{tabular}
\end{center}
\vskip .5cm
\begin{center}
{Table IIa- Average values of kinematic quantities and 
correction factors - H1 $k_T> 3.5 GeV$}
\end{center}

\bigskip\bigskip

\begin{center}
\begin{tabular}{|c|c|c|c|c|c|} \hline 
 x & $\sigma$ & $Q^2$ & $E_{jet}$ & $k_T$ & Corr. Factor (.$10^{-3}$)\\
\hline\hline
 0.00036 & 27.5 & 18.0 & 35.9 & 5.5 & 0.108 \\
 0.00073 & 126. & 27.0 & 36.5 & 5.8 & 0.695 *\\
 0.0012 & 132. & 32.2 & 37.9 & 6.3 & 0.895 \\
 0.0017 & 96. & 34.8 & 39.3 & 6.5 & 0.979 \\
 0.0024 & 55. & 40.1 & 39.4 & 6.7 & 0.870 \\
 0.0035 & 28. & 48.2 & 39.6 & 7.2 & 0.696 \\
\hline\hline
\end{tabular}
\end{center}
\vskip .5cm
\begin{center}
{Table IIb- Average values of kinematic quantities and 
correction factors - H1 $k_T> 5 GeV$}
\end{center}
\bigskip\bigskip

\begin{center}
\begin{tabular}{|c|c|c|c|c|c|} \hline 
 x & $\sigma$ & $Q^2$ & $E_{jet}$ & $k_T$ & Corr. Factor (.$10^{-3}$)\\
\hline\hline
 0.0006 & 114.0 & 28.0 & 36.5 & 6.3 & 0.304 * \\
 0.0011 & 96.2 & 39.0 & 38.0 & 6.9 & 0.656 \\
 0.0019 & 77.8 & 50.7 & 39.9 & 7.6 & 0.966 \\
 0.0033 & 34.4 & 75.6 & 43.8 & 8.7 & 0.996 \\
 0.006 & 14.1 & 113.6 & 49.6 & 10.4 & 0.995 \\
 0.01 & 6.53 & 176.4 & 58.5 & 12.9 & 0.896 *\\
 0.018 & 2.65 & 244.7 & 67.3 & 15.1 & 0.653 *\\
 0.031 & 0.65 & 366.8 & 78.8 & 18.8 & 0.373 *\\
\hline\hline
\end{tabular}
\end{center}
\vskip .5cm
\begin{center}
{Table IIc- Average values of kinematic quantities and 
correction factors - ZEUS $k_T> 5 GeV$}
\end{center}

The correction factors are given in Table {\bf IIa} for H1 ($k_T > 3.5$ GeV), 
Table {\bf 
IIb} for H1 ($k_T > 
5$ GeV), and
Table {\bf IIc} for ZEUS bins together with the value of the bin centers 
determined with 
the full Monte-Carlo
simulation~\cite{ariadne}, and the experimental values of the 
cross-sections\footnote{ 
Note that we did 
not use the
full Monte-Carlo to get the correction factors in order to avoid any strong 
model 
dependence as
these factors are only due to kinematic effects. It is however more difficult to 
use a 
toy Monte 
Carlo
to get accurate values for the bin centers, and this is why we used a full Monte 
Carlo 
for this sake.
However, the dependence of the theoretical cross-section on the bin centers is
minimized by our specific choice of kinematic variables (see formula (7)). }. We 
note 
that the 
correction
factors are quite different from one $x$-bin to an other
and much less than one (in $10^{-3}$ units), explicitely showing that the 
experimental
cuts play an important role in the cross-section measurement, and that these 
factors are 
compulsory
to be taken into account if we want to get a direct comparison with the 
theoretical 
cross-sections. We 
also note that the correction factors 
are very much different from one another at very low $x$, showing that the
acceptance of these bins is quite low. This is also why it is not so
easy to be able to get a correct value of the measured cross-section after
cuts in those bins.
We also get the same order of magnitude for the correction factors for the H1 
and ZEUS 
experiments 
because 
the experimental cuts are quite similar. 
The differences between both experiments are
due mainly to the fact that the range in $x$ and $Q^2$ is
much lower for H1 than for ZEUS (the reference bin for H1 goes to lower $Q^2$ 
compared 
to ZEUS). 
\bigskip

{\bf 4. Fits}

\bigskip

Using the kinematic correction factors determined as described in the previous 
section, we perform a 
fit to the H1 and ZEUS data with only two free parameters. these are the {\it 
effective} 
strong 
coupling
constant in LO BFKL formulae $\bar{\alpha}$ corresponding to the {\it effective} 
Lipatov 
intercept
$\alpha_P= 1+4 \log 2 \bar{\alpha} N_C/\pi$, and the cross-section 
normalisation. The 
obtained values
of the parameters and the $\chi^2$ of the fit are given in Table {\bf III} for a 
fit to 
the H1 and ZEUS 
data
separately, and then to the H1 + ZEUS data together. Note that 
one H1 point at $k_T > 5$ GeV ($7.3$ 10$^{-4}$), and four ZEUS points
($x=4.$ 10$^{-4}$, and the three highest-x points), were not taken into account 
in the 
fit and are 
distinguished in Tables II with a 
star. We will discuss this selection in a little while.

\begin{center}
\begin{tabular}{|c|c|c|c|c|c|c|} \hline
 fit & $\bar{\alpha}$ & $\alpha_P$ & Norm. & $\chi^2 (/dof)$ \\ 
\hline\hline
 H1 & 0.17 $\pm$ 0.02 $\pm$ 0.01 & 1.44 $\pm$ 0.05 $\pm$ 0.025 & 29.4 $\pm$ 4.8
 $\pm$ 
5.2 & 5.7 (/9)\\
 ZEUS & 0.20 $\pm$ 0.02 $\pm$ 0.01 & 1.52 $\pm$ 0.05 $\pm$ 0.025 & 26.4 $\pm$
 3.9 $\pm$ 
4.7 & 2.0 (/2)\\
 H1+ZEUS & 0.16 $\pm$ 0.01 $\pm$ 0.01 & 1.43 $\pm$ 0.025 $\pm$ 0.025 & 30.7
 $\pm$ 2.9 
$\pm$ 3.5 & 12.0 (/13)\\
\hline
\end{tabular}
\end{center}
\vskip .5cm
\begin{center}
{Table III- Fit results}
\end{center}

The $\chi^2$ of the fits have been calculated using statistical error only and 
are very 
satisfactory
(about $0.6 \ per \ point$ for H1 data, and $1. \ per \ point$ for ZEUS data).
We give both statistical and systematic errors for the fit parameters. 
The values of the Lipatov intercept are close to one another and compatible
within errors for the H1 and ZEUS sets of data, and indicate a preferable medium 
value 
($\alpha_P=1.4-1.5$). We also notice that the ZEUS data have
the tendency to favour a higher exponent, but the number of data points
used in the fit is much smaller than for H1, and the H1 data are also at
lower $x$. The normalisation
is also compatible between ZEUS and H1. The fit results are shown in Figure 1 
and 
compared with
the H1 and ZEUS measurements.

Let us discuss our selection criterium for the fits. Both lowest $x$ points for 
H1 and 
ZEUS
 show large correction factors but only the lowest $x$ point for  ZEUS lies a 
bit above 
the prediction, 
which shows the relevance of the correction factors we determined. On the other 
hand,  
the three 
highest $x$ points
for ZEUS cannot be described by a BFKL fit probably because the $x$-value is too
high ($x> 10^{-2}$). Consider now the second lower $x$ point at $k_T
> 5$ GeV for the H1 experiment that we suppressed from the fit (see Table {\bf 
IIb}). If 
we include it in the 
fit the $\chi^2$ value goes from 5.7 to 32, which
is due to the small statistical error of this data point (the systematic
error is on the contrary very large). By comparison, 
including the lowest x point for ZEUS changes the $\chi^2$ from 2.0 to 7.9. In 
the same 
way,
 including the highest x points still increases the $\chi^2$ to 67.4, showing 
clearly
that these highest x  points cannot be described using the BFKL formalism. It is 
interesting to note 
that all similar discrepancies appear also in other types of fitting procedures, 
e.g. 
in Ref. 
\cite{kw99}. 

\bigskip

{\bf 5. Comparison with other processes}

\bigskip

The final result of our new determination of the effective pomeron
intercept is $\alpha_P=1.43 \pm 0.025$ (stat.) $\pm 0.025$ (syst.).
This high value of the intercept leads to the following remarks.
Our analysis confirms the trend observed using DGLAP based Monte-Carlo
\cite{po99} which have difficulties to reproduce the forward jet cross-section
due to a low effective pomeron intercept when both $k_T^2$ and $Q^2$ scales
are of the same order. 

On the other hand, our method allows a direct comparison of the intercept 
values with those obtained in other experimental processes, i.e.
$\gamma^* \gamma^*$ cross-sections at LEP \cite{ro99}, jet-jet cross-sections
at Tevatron at large rapidity intervals \cite{D0}, $F_2$ and $F_2^D$ proton
structure function measurements \cite{ab00,na96,mu98}. Let us first consider the
known determinations of the effective intercepts in $F_2$ and $F_2^D$
measurements at HERA \cite{h1ZEUS}.
It is known that the effective intercept determined in these measurements is
rather low\footnote{It is interesting to note that the ``hard'' Pomeron 
intercept obtained within the framework of two-Pomeron models\cite{la99}
fits with our determination. However our parametrization (\ref{dsigma}) 
corresponds to only one Pomeron.}(1.2-1.3). This is the reason why these data 
can be both described
by a DGLAP or a BFKL-LO fit \footnote{ Note that in the BFKL 
descriptions of these data \cite{na96,mu98}, the effective intercept is
taken to be constant, while 
the $Q^2$ dependence comes from the BFKL integration (see  for instance
formula (\ref{dsigma}))}.

\begin{figure}
\begin{center}
\centerline{\psfig{figure=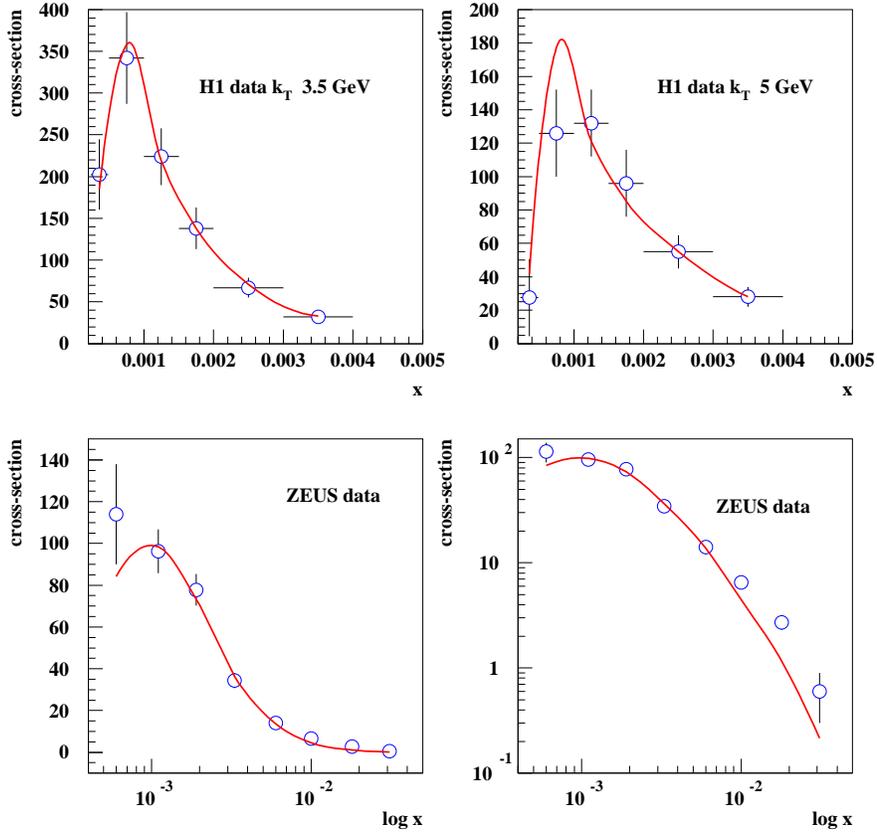,height=5.in}}
\end{center}
\caption{The H1 data ($k_T > 3.5$ GeV, $k_T > 5$ GeV), and the ZEUS data are
compared with the result of the fit. ZEUS data are also displayed in 
logarithmic scales in vertical coordinates to show the discrepancy at high
$x$ values.}
\end{figure}

Now let us consider processes initiated by two hard probes which allow 
a more direct comparison between experiments and BFKL predictions. 
These processes suppress DGLAP evolution by selecting events
with comparable hard scales for both hard probes. 
Recent data on $\gamma^* \gamma^*$ cross-section measurements at LEP
\cite{opall3} lead to a BFKL description with a low effective intercept 
compatible with the one of $F_2$ and $F_2^D$ at HERA 
($\alpha_P$=1.2-1.3 \cite{ro99}) \footnote{The statistics for these data  
is still very low. L3 and OPAL Collaborations have released the cuts used
to enhance BFKL effects to get more statistics \cite{opall3,ro99}. 
These data can be 
both described by BFKL and DGLAP evolution equations.}.  
The fact that similar values of the intercepts are found could be interpreted
by sizeable higher order corrections to BFKL equation. 
On the other hand, 
it is interesting to note that our
result based on forward jet measurement at HERA obtained in comparable
$Q^2$ ($Q^2 \sim 10$ GeV$^2$) and rapidity 
($Y \sim$ 3-4) domains is quite different. The value of the intercept is
significantly higher.

It is also fruitful to compare our results with the effective intercept 
we obtain from recent preliminary dijet data obtained by the
D0 Collaboration at Tevatron \cite{D0}. The measurement consists
in the ratio $R=\sigma_{1800}/ \sigma_{630}$ where $\sigma$ is the dijet 
cross-section at large rapidity interval $Y \sim \Delta \eta$ 
for two center-of-mass energies
(630 and 1800 GeV), $\Delta \eta_{1800}=4.6$, $\Delta \eta_{630}=2.4.$
The experimental measurement is $R=2.9 \pm 0.3$ (stat.) $\pm 0.3$ (syst.).
Using the Mueller-Navelet formula \cite{mu86}, this measurement allows us to get 
a value 
of the effective intercept for this
process \footnote{Formula (\ref{muelnav}) is obtained after integration over 
the jet tranverse energies at 630 and 1800 GeV, $E_{T_1}$, $E_{T_2}$. 
We note that a non integrated formula
shows a sizeable dependence on $E_{T_1}/E_{T_2}$, which could be confronted
with experiment \cite{conftevatron}.} 
\begin{eqnarray}
R=\frac{\int_{\frac 12- i\infty}^{\frac 12+ i\infty} \frac{d \gamma}{2i \pi
\gamma (1-\gamma)}
e^{\epsilon (\gamma,0) \Delta \eta_{1800}}} 
{\int_{\frac 12- i\infty}^{\frac 12+ i\infty} \frac{d \gamma}
{2i \pi \gamma (1-\gamma)}
e^{\epsilon (\gamma,0) \Delta \eta_{630}}}. 
\label{muelnav}
\end{eqnarray}
We get $\alpha_P$=1.65 $\pm$ 0.05 (stat.) $\pm$ 0.05 (syst.), in agreement with
the value obtained by D0 using a saddle-point approximation \cite{D0}.
This intercept is higher than the one obtained in the forward jet study.

The question arises
to interpret the different values of the  effective intercept. It could 
reasonably
come from the differences in higher 
order QCD corrections for the BFKL kernel and/or in the impact factors
depending on the initial probes ($\gamma^*$
vs. jets).
In order to evaluate the approximate size of the higher order BFKL corrections,
we will use their description in terms of rapidity veto effects \cite{nlo}.
In formula (\ref{dsigma}), we make the following replacement 
\begin{eqnarray}
 \exp (\epsilon (\gamma,0) Y) \rightarrow 
\Sigma _{n=0}^{\infty}~ \theta (Y-(n+1)b)~ \frac{\left[ \epsilon(\gamma,0) 
~(Y-(n+1)b)
\right]^n}
{\Gamma(n+1)}~. 
\end{eqnarray}
The Heaviside function $\theta$ ensures that a BFKL ladder of
$n$ gluons occupies $(n+1)b$ rapidity interval where $b$ parametrises the 
strength of NLO
BFKL corrections. The value of the leading order intercept is fixed to 
$\alpha_p=1.75 (\alpha_S(Q^2=10)=0.28)$, where $Q^2=10$ GeV$^2$ is
inside the average range of $Q^2$ in the forward jet measurement. 
The fitted value of the $b$ parameter obtained
using the forward jet data is found to be 1.28 $\pm$ 0.08 (stat.) $\pm$ 0.02 
(syst.).
Imposing the same value of $\alpha_P$ with Tevatron data gives
$b$=0.21 $\pm$ 0.11 (stat.) $\pm$ 0.11 (syst.). Note that the theoretical value 
of $b$ for the NLO BFKL kernel is expected to be of the order 2.4, which is also 
compatible with 
the result obtained
for the $\gamma^* \gamma^*$ cross-section. A contribution from the NLO impact
factors is not yet known, and could perhaps explain the different values of $b$.

\bigskip

{\bf 6. Conclusions}

\bigskip

To summarize our results, using a new method to disantangle the effects of the
kinematic cuts from the genuine dynamical values of the forward jet 
cross-sections
at HERA, we find that the effective pomeron intercept is $\alpha_P=1.43 \pm 
0.025$
(stat.) $\pm 0.025$ (syst.). It is much higher than the soft pomeron intercept,
and, among those determined in hard processes,  it is intermediate 
between $\gamma^* \gamma^*$
interactions at LEP and dijet productions with large rapidity intervals at 
Tevatron.

Looking for an interpretation of our results in terms of higher order BFKL 
corrections expressed by rapidity gap vetoes $b$ between emitted gluons, we find 
a value of $b=$1.3, which is sizeable but less than the theoretically predicted
\cite{li98}
value for the NLO BFKL kernel ($b=$2.4). The observed dependence in the process
deserves further more precise studies \cite{lipatov}.

Last but not least, the derivation of the correction factors given in Table II
is independent of the theoretical input and could be used to test any model
suitable for the jet cross-section.

\bigskip

{\bf Acknowledgments}

\bigskip

We would like to thank Lev Lipatov for his fruitful remarks and suggestions.
One of us (J.G.C.) acknowledges supports by CONACyT. 

%%%%%%%%%%%%%% 

\end{document}